\documentclass{article}
\usepackage{amssymb}

\title{A use of the central limit theorem to obtain a classical limit for the center-of-mass tomogram}

\author{G.G. Amosov and V.I. Man'ko}

\begin {document}

\maketitle

\begin{abstract}

We investigate the dependence of the center-of-mass tomogram of a system with many degrees of freedom $N$ on the Planck constant $\hbar $.
It is shown that to use the central limit theorem under taking the limit $N\to +\infty $ one should keep the energy of the system to be constant.
In the case, the resulting distribution is Gaussian if the initial distribution is a product of independent excited states
of a quantum oscillator or even and odd coherent states either. Then, if one turns the Planck constant $\hbar \to 0$ we get $\delta $-function 
associated with the distribution concentrated in zero with the probability equal to one.

\end{abstract}

\section {Introduction}

In \cite {A05} we studied the center-of-mass tomogram \cite {Ar04, Ar05} in the limiting case of many degrees of freedom $N\to +\infty$. It was shown that 
the final distribution tends to Gaussian if the state of the initial system is a product of excited states of a quantum oscillator and some additional conditions
are satisfied. The conditions obtained are due to the central limit theorem. Nevertheless the uniteless case was considered (the Planck constant $\hbar =1$). 
In the present paper we investigate the classical limit $\hbar \to 0$ for the center-of-mass tomogram. Initially we were inspired by the study of 
taking the Ehrenfest limit ($\hbar \to 0$ under preserving the mean energy) discussed in \cite {M05} in the context of quantum tomograms. 
Nevertheless, the results obtained are not of the same sort as in \cite {M05}.

Denote $\mathfrak {S}(H)$ the set of quantum states (positive unite-trace operators) in a Hilbert space $H$.
We shall use the well-known formalism of quantum probability theory (see, f.e. \cite {Hol}) in which given an observable $\hat x$ in $H$  and 
a quantum state $\hat \rho\in  \mathfrak {S}(H)$ one can define the probability distribution
\begin {equation}\label {dist}
P(\hat x\in \Omega )=Tr(\hat \rho \hat E_{\hat x}(\Omega )),
\end {equation}
where $\hat E_{\hat x}$ is a spectral projection-valued measure associated with $\hat x$ such that $\hat x=\int \limits Xd\hat E_{\hat x}((-\infty ,X])$ and
$\Omega $ is an arbitrary Borel subset of the real line $\mathbb R$. We shall say that the observable $\hat x$ has a probability distribution
(\ref {dist}) in the state $\hat \rho $. We denote $\mathbb {E}(\hat x)$ and $Var(\hat x)$ the expectation and the variance of a random
value with the probability distribution (\ref {dist}). The sense of (\ref {dist}) is the statistical readings of the instrument measuring $\hat x$ in the state $\hat \rho $.
Different observables $\hat x_1,\dots ,\hat x_n$ are said to be independent in the state $\hat \rho $ if they are pairwise commuting and the common
distributions of $\hat x_1,\dots ,\hat x_n$ determined by the formula
$$
P(\hat x_1\in \Omega _1,\dots ,\hat x_N\in \Omega _N)=Tr(\hat \rho \hat E_{\hat x_1}(\Omega _1)\dots \hat E_{\hat x_N}(\Omega _N))
$$
are factorizable in the sense that
$$
P(\hat x_1\in \Omega _1,\dots ,\hat x_N\in \Omega _N)=\prod _{i=1}^NP(\hat x_i\in \Omega _i).
$$

\section {The center-of-mass tomography}

Let $H=\oplus _{i=1}^NH_i$ be a Weyl-quantized N-mode systems with the independent canonical observables $(\hat q_1,\hat p_1),\dots ,(\hat q_N,\hat p_N)$.
Following to \cite {Ar04, Ar05} we define the center-of-mass tomogram $\omega _{cm}$ of a quantum state $\hat \rho \in \mathfrak {S}(H)$ 
as a function of $2N+1$ parameters $\overline \mu =(\mu 1,\dots ,\mu _N)^T,\  \overline \nu =(\nu _1,\dots ,\nu _N)^T$ and $X\in {\mathbb R}$
by the formula 
$$
\omega _{cm}(X,\overline \mu ,\overline \nu)=Tr(\hat \rho \delta (X-\overline \mu \overline {\hat q} -\overline \nu \overline {\hat p})),
$$ 
where $\overline \mu \overline {\hat q}=\sum \limits _{i=1}^N\mu _i\hat q_i,\  \overline \nu \overline {\hat p}=\sum \limits _{i=1}^N\nu _i\hat p_i$, respectively.
Since the center-of-mass tomogram $\omega _{cm}$ is set one can restore the state $\hat \rho $ as follows
\begin{equation}\label {state}
\hat \rho =\frac {1}{(2\pi )^N}\int \exp [i(X-\overline \mu \overline {\hat q}-\overline \nu \overline {\hat p})]\omega _{cm}(X,\overline \mu ,\overline \nu)dXd\overline \mu d\overline \nu.
\end{equation}

Suppose that $\hat \rho $ is a separable pure state such that $\hat \rho =\prod _{i=1}^N\hat \rho _i,\ \hat \rho _i\in \mathfrak {S}(H_i)$. Introducing the symplectic 
tomograms $\omega _i$ of the states $\hat \rho _i$ by the formula
$$
\omega _i(X,\mu _i,\nu _i)= Tr(\hat \rho _i\delta (X-\mu _i\hat q_i-\nu _i\hat p_i))
$$
we obtain the center-of-mass tomogram of $\hat \rho $ in the following form
\begin{equation}\label{gig}
\omega _{cm}(X,\overline \mu ,\overline \nu)=\int \delta (X-\sum \limits _{j=1}^NY_j)\prod _{j=1}^N\omega _j(Y_j,\mu _j,\nu _j)d\overline Y.
\end{equation}

Although the expression (\ref {gig}) seems to be complicated, it is possible to obtain its limiting form if $N\to +\infty$ under some conditions (see \cite {A05}).
In fact, let us involve the observables 
\begin{equation}\label{x}
\hat x_i=\hat x_i(\mu _i,\nu _i)=\mu _i\hat q_i+\nu _i\hat p_i,\ 
1\le i\le N, 
\end{equation}
which are independent in the state (\ref {state}). Put 
\begin{equation}\label {s}
\hat s_N=\hat s_N(\overline \mu, \overline \nu)=\sum \limits _{i=1}^N\hat x_i.
\end{equation}
Then, the center-of-mass tomogram (\ref {tomogr}) can be considered as a probability distribution of $\hat s_N$ in the separable pure state $\hat \rho$.
The classical condition of the central limit theorem (the Lyapunov theorem, \cite {Lyap}, P. 371) reads
\begin{equation}\label{cond}
S_N\equiv \frac {\sum \limits _{j=1}^N\mathbb {E}(|\hat x_j|^3)}{(Var(\hat s_N))^{3/2}}\to 0,\ N\to +\infty .
\end{equation}
If (\ref {cond}) is satisfied the limiting probability distribution of (\ref {tomogr}) is asymptotically Gaussian for each fixed $\mu _i,\nu _i,\ 1\le i\le N$.
Our goal is to reveal what kind of conditions we need. In the next section we shall apply to (\ref {gig}) the central limit theorem for two different cases,
where all $\rho _i$ are excited states of a quantum oscillator and even or odd coherent states either.

\section {Applying the central limit theorem}

\subsection {Excited states of a quantum oscillator}

Let $\hat \rho $ be compiled from $N$ excited states of a quantum oscillator such that its wave function in the coordinate representation is
$$
<\overline X\ |\ \hat \rho >=\prod _{i=1}^N\frac {1}{(\pi \hbar )^{1/4}}\frac {1}{\sqrt {2^{n_i}n_i!}}H_{n_i}\left (\frac {X_i}{\sqrt {\hbar }}\right )\exp\left (-\frac {X_i^2}{2\hbar }\right ),
$$
where $\overline X=(X_1,\dots ,X_N)^T$, $H_n$ denotes the Hermite polynomial and we put $m=\Omega =1$ for the mass as well as the frequency.
In the case, taking into account that we do not put $\hbar =1$ like it was done in \cite {Ar04, Ar05}) we get
$$
\omega _{cm}(X,\overline \mu ,\overline \nu )=\int \delta (X-\sum \limits _{i=1}^NY_j)\times 
$$
\begin{equation}\label{tomogr}
\prod _{j=1}^N\frac {1}{\sqrt {\pi \hbar (\mu _j^2+\nu _j^2)}}\frac {1}{2^{n_j}n_j!}
H_{n_j}^2\left (\frac {Y_j}{\sqrt {\hbar (\mu _j^2+\nu _j^2)}}\right )\exp\left (-\frac {Y_j^2}{\hbar (\mu _j^2+\nu _j^2)} \right )d\overline Y.
\end{equation}

Notice that the energy of a system in the state (\ref {state}) equals
\begin{equation}\label {energy}
E=\hbar \left (\frac {N}{2}+\sum \limits _{i=1}^Nn_i \right ).
\end {equation}
Thus, the Ehrenfest limit suppose that we keep the value (\ref {energy}) to be constant.

The definitions (\ref {x}) and (\ref {s}) imply that
$$
\mathbb {E}(\hat x_i)=\mathbb {E}(\hat s_N)=0,
$$ 
$$
Var(\hat x_i)=\hbar (\mu _i^2+\nu _i^2)(\frac {1}{2}+n_i),
$$
$$
Var(\hat s_N)=\hbar \left (\sum \limits _{i=1}^N(\mu _i^2+\nu _i^2)(\frac {1}{2}+n_i)\right ),
$$
and
$$
\mathbb {E}(|\hat x_j|^3)=2\int \limits _0^{+\infty }X^3\frac {1}{\sqrt {\pi \hbar (\mu _j^2+\nu _j^2)}}\frac {1}{2^{n_j}n_j!}
H_{n_j}^2\left (\frac {X_j}{\sqrt {\hbar (\mu _j^2+\nu _j^2)}}\right )\times 
$$
$$
\exp\left (-\frac {X_j^2}{\hbar (\mu _j^2+\nu _j^2)} \right )dX\le Cn_j^{3/2}(\hbar (\mu _j^2+\nu _j^2))^{3/2},
$$
where the constant $C$ doesn't depend on $\mu _j,\nu _j$ and $n_j$.

Because the homogeneity of tomograms, to reconstruct the tomogram, it is sufficiently to consider $\mu _i,\nu _i$ delimited below from zero and bounded from above uniformly in $i$ such that
$r<\mu _i^2+\nu _i^2<R$ uniformly in $i$. It results in the following inequalities
\begin{equation}\label{est}
rE\le Var(\hat s_N)\le RE,
\end{equation}
where the energy $E$ is determined by (\ref {energy}). 

One can see that $S_N$ doesn't depend on $\hbar $ and can be estimated as 
$$
S_N\le A\frac {\sum \limits _{i=1}^Nn_i^{3/2}}{\left (\frac {N}{2}+\sum \limits _{i=1}^Nn_i\right )^{3/2}}
$$
with the constant $A$ depending only on the energy of the system $E$ and the constants of (\ref {est}). 
It follows that if all the numbers $n_i$ are uniformly bounded from above, then $S_N\to 0$ while $N\to +\infty $
and the probability distributions determined by the central-of-mass tomogram (\ref {tomogr}) are asymptotically 
Gaussian with the zero mean and the variance equal to 
$$
 \sigma ^2_N=\hbar \left (\sum \limits _{i=1}^N(\mu _i^2+\nu _i^2)(\frac {1}{2}+n_i)\right )
 $$
with the borders caused by (\ref {est})
$$
rE\le \sigma ^2_N\le RE.
$$

Now fix the number $N$ under which the distribution of the center-of-mass tomogram $\omega _{cm}$ is almost Gaussian.
Then, the limt $\hbar \to 0$ results in $\sigma ^2_N\to 0$ and we get $\omega _{cm}\to \delta (X)$ for all the pairs $(\mu_i, 
\nu_i)$ distinguished from zero. Thus, we obtain the probability distribution concentrated in zero with the probability one.

\subsection {Even and odd coherent states}

Given a complex parameter $\alpha $ one can define the coherent state $|\alpha >$ as
$$
|\alpha >=\exp(-\frac {|\alpha |^2}{2})\sum \limits _{n=0}^{+\infty }\frac {\alpha ^n}{\sqrt {n!}}|n>,
$$
where we denoted $|n>$ the excited states of a quantum oscillator.
The wave function of $|\alpha >$ in the coordinate representation reads
$$
<X|\alpha >=\frac {1}{(\pi \hbar )^{1/4}}\exp(-\frac {|\alpha |^2}{2}-\frac {X^2}{2\hbar }+\frac {\sqrt \alpha X}{\sqrt \hbar }-\frac {\alpha ^2}{2}).
$$
In \cite {DMM} it was proposed to consider superpositions of two coherent states $|\alpha >$ and $|-\alpha >$.
In this framework, the even coherent state is defined by
$$
|\alpha _+>=N_+(|\alpha >+|-\alpha>),\ N_+=\frac {\exp(\frac {|\alpha |^2}{2})}{2\sqrt {\cosh|\alpha |^2}},
$$
while the odd coherent state is given by
$$
|\alpha _->=N_-(|\alpha >-|-\alpha>),\ N_-=\frac {\exp(\frac {|\alpha |^2}{2})}{2\sqrt {\sinh|\alpha |^2}},
$$

The symplectic quantum tomograms $\omega _{\pm}$ associated with the even and odd coherent states $|\alpha _{\pm}>$
are determined as follows (\cite {M})
$$
\omega _{\pm}(X,\mu ,\nu)=\frac {N_{\pm }}{\sqrt {\pi \hbar (\mu ^2+\nu ^2)}}\exp \left [-\frac {1}{2}(\alpha +\alpha ^*)^2-\frac {X^2}{\hbar (\mu ^2+\nu ^2)}\right .
$$
\begin{equation}\label{coherent}
\left . +\nu \left (\frac {\alpha ^2}{\nu -i\mu }+\frac {\alpha ^{*2}}{\nu +i\mu}\right )\right ] 
\left |\exp\left (\frac {i\sqrt 2\alpha X}{\hbar (i\mu -\nu )}\right )\pm \exp\left (-\frac {i\sqrt 2\alpha X}{\hbar (i\mu -\nu)}\right )\right |^2 
\end{equation}
Notice that the probability distributions of (\ref {coherent}) are not Gaussian. 

Now let us consider the separable pure state $\hat \rho =\prod _{i=1}^N|\alpha _{i\#}><\alpha _{i\#}|$, where $\alpha _i\in {\mathbb C}$
and $\#$ equals $+$ or $-$. We shall denote $N_{i\pm}$ the normalization constant for the state $|\alpha _i\#>$ with a choice of
the index depending on $\# =+$ or $-$ either. Taking into account that for the observables (\ref {x}) the following formulae hold,
$$
<\alpha _i|\hat x_i|\alpha _i>=\sqrt {2\hbar }(Re(\alpha _i)\mu _i +Im(\alpha _i)\nu _i),
$$
$$
<-\alpha _i|\hat x_i|\alpha _i>=i\sqrt {2\hbar } e^{-2|\alpha _i|^2}(Im(\alpha _i)\mu _i+Re(\alpha _i)\nu _i),  
$$
$$
<\alpha _i|\hat x_i^2|\alpha _i>=\frac {\hbar }{2}(\mu _i^2+\nu _i^2)+2\hbar (Re(\alpha _i)\mu _i +Im(\alpha _i)\nu _i)^2,
$$ 
$$
<-\alpha _i|\hat x_i^2|\alpha _i>=e^{-2|\alpha _i|^2}\left (\frac {\hbar }{2}(\mu_i ^2+\nu_i ^2)-2\hbar (Im(\alpha _i)\mu _i+Re(\alpha _i)\nu _i)^2\right ),
$$
we get for the expectations and variances in the states $\hat \rho _i$
$$
{\mathbb E}(\hat x_i)=0,
$$
$$
Var(\hat x_i)=N_{i\pm }\hbar \left ((1+e^{-2|\alpha _i|^2})(\mu _i^2+\nu _i^2)+\right .
$$
$$
\left .4(Re(\alpha _i)\mu _i +Im(\alpha _i)\nu _i)^2\mp 4e^{-2|\alpha _i|^2}(Im(\alpha _i)\mu _i+Re(\alpha _i)\nu _i)^2\right ).
$$

Suppose that $\mu _i^2+\nu _i^2=\rho $. Then, a rough estimation gives us
$$
Var(\hat x_i)\ge N_{i\pm }\hbar \rho (1+e^{-2|\alpha _i|^2}(1-4|\alpha _i|^2))\ge C_1N_{i\pm}\hbar ,
$$
$$
{\mathbb E}(|x_i|^3)\le C_2\hbar ^{3/2}|\alpha _i|^3,
$$
$$
C_3<N_{i+}<C_4,\ \frac {C_5}{\sqrt {|\alpha |}}<N_{i-}-<C_6,
$$
where the constants $C_i,\ 1\le i\le 6,$ do not depend on $\alpha _i,\mu _i,\nu _i$.

Similarly to the previous section, let us claim that $r\le \mu _i^2+\nu _i^2<R$ and $|\alpha _i|\le A$ uniformly in $i$. Then, we obtain the quantity (\ref {cond}),
$$
S_N\le B\frac {\sum \limits _{i=1}^N|\alpha _i|^3}{(\sum \limits _{i=1}^NN_{i\#})^{3/2}}\to 0,\ N\to +\infty ,
$$
where the constant $B$ doesn't depend on $\hbar $.
Thus, the distributions of $\omega _{cm}$ are asymptotically Gaussian with the zero mean and the variance equal to
$$
\sigma ^2_N=\hbar \sum \limits _{i=1}^NN_{i\pm } \left ((1+e^{-2|\alpha _i|^2})(\mu _i^2+\nu _i^2)+\right .
$$
$$
\left .4(Re(\alpha _i)\mu _i +Im(\alpha _i)\nu _i)^2\mp 4e^{-2|\alpha _i|^2}(Im(\alpha _i)\mu _i+Re(\alpha _i)\nu _i)^2\right ).
$$
Taking the limit $\hbar \to 0$ we conclude that $\sigma ^2_N\to 0$ and the final distribution is concentrated in zero with the probability equal to one.

\section {Conclusion} 

We considered the behavior of the center-of-mass tomogram $\omega _{cm}$ in the limit case of many degrees of freedom $N$. The result
of \cite {A05} is concretized by taking into account the dependence of $\omega _{cm}$ on the Planck constant $\hbar$. 
We considered excited states of a quantum oscillator of \cite {A05} and a new example of even and odd coherent states.
It is shown that
the limiting distribution is Gaussian under the claim that we keep the energy of the system to be constant. Then, if
$\hbar \to 0$ we get the $\delta $-function giving the distribution concentrated in zero with the probability one.

\section*{Acknowledgments} 

This work was partially supported by the Russian Foundation for
Basic Research under Projects Nos. 07-02-00598 and 08-02-90300-Viet (G.G.A. and V.I. M.)
and by the grant ADTP "Evaluating the scientific potential of high school" Pr. Nr. 2.1.1/1662 (G.G.A.).

\end {document}